\documentclass[twocolumn,amsmath,amssymb,pre]{revtex4}
\usepackage{mhchem}
\usepackage{times,mathptmx}
\usepackage{amssymb}
\usepackage{times}
\usepackage{balance}
\usepackage{graphicx} 
\usepackage[format=plain,justification=raggedright,singlelinecheck=false,font=small,labelfont=bf,labelsep=space]{caption}

\begin{document}

\newcommand{\degc}{\ensuremath{^{\circ}}\mathrm{C}}

\def\cp{C_{\textrm{p}}}            

\makeatletter

\title{Phase separation dynamics in colloid-polymer mixtures: the effect of interaction range}

\author{Isla Zhang}
\affiliation{School of Chemistry, University of Bristol, Bristol, BS8 1TS, UK.}

\author{C. Patrick Royall}
\affiliation{HH Wills Physics Laboratory, University of Bristol, Tyndall Avenue, Bristol BS8 1TL, UK, School of Chemistry, University of Bristol, Bristol BS8 1TS, UK,
Centre for Nanoscience and Quantum Information
Tyndall Avenue, Bristol BS8 1FD, UK.}

\author{Paul Bartlett}
\affiliation{School of Chemistry, University of Bristol, Bristol, BS8 1TS, UK.}

\author{Malcolm A. Faers}
\affiliation{Bayer CropScience AG, 40789, Monheim am Rhein, Germany.}

\begin{abstract}
Colloid-polymer mixtures may undergo either fluid-fluid phase separation or gelation. This depends on the depth of the quench (polymer concentration) and polymer-colloid size ratio.  We present a real-space study of dynamics in phase separating colloid-polymer mixtures with medium- to long-range attractions (polymer-colloid size ratio $q_{\textrm{R}}=0.45-0.89$), with the aim of understanding the mechanism of gelation as the range of the attraction is changed. In contrast to previous studies of short-range attractive systems, where gelation occurs shortly after crossing the equilibrium phase boundary, we find a substantial region of fluid-fluid phase separation. On deeper quenches the system undergoes a continuous crossover to gel formation.  
We identify two regimes, `classical' phase separation, where single particle relaxation is faster than the dynamics of phase separation, and `viscoelastic' phase separation, where demixing is slowed down appreciably due to slow dynamics in the colloid-rich phase. Particles at the surface of the strands of the network exhibit significantly greater mobility than those buried inside the gel strand which presents a method for coarsening.\end{abstract}

\maketitle

\section{Introduction}

Dynamical arrest in soft matter can generally be divided into two classes, glasses and gels. Arrest is exhibited by a wide variety of systems,
including those which can be treated as spheres with attractions which include colloid-polymer mixtures \cite{poon2002} and protein solutions \cite{cardinaux2007, gibaud2009}. Dynamical arrest occurs upon either a global increase in density (vitrification) or arrested phase separation (gelation). In the latter case, phase separation is suppressed as the system undergoes arrest to form a non-equilibrium gel state, which we define as a percolating assembly of particles \cite{lu2008,manleyWyss2005} which do not relax on the experimental timescale.


In colloid-polymer mixtures, addition of sufficient polymer causes demixing into two or more phases \cite{asakura1954,vrij1976}. In a one-component description, the degrees of freedom of the polymers are formally integrated out, leading to an effective attraction between the colloidal particles whose range is fixed by the ratio of the polymer and colloid diameters $q_\textrm{R}$ \cite{dijkstra1999} and whose strength is set by the polymer concentration. To a reasonable approximation, polymer concentration plays the role of inverse temperature. The topology of the phase diagram is well-known: short-range interactions lead to only fluid-crystal phase coexistence, along with a metastable fluid-fluid binodal. However as the interaction range increases to $q_{\textrm{R}}\approx 0.3$, fluid-fluid phase separation produces a stable colloidal liquid and colloidal gas \cite{lekkerkerker1992,poon2002,ilett1995}.

\begin{figure}[h]
\centering
\includegraphics[width=80mm]{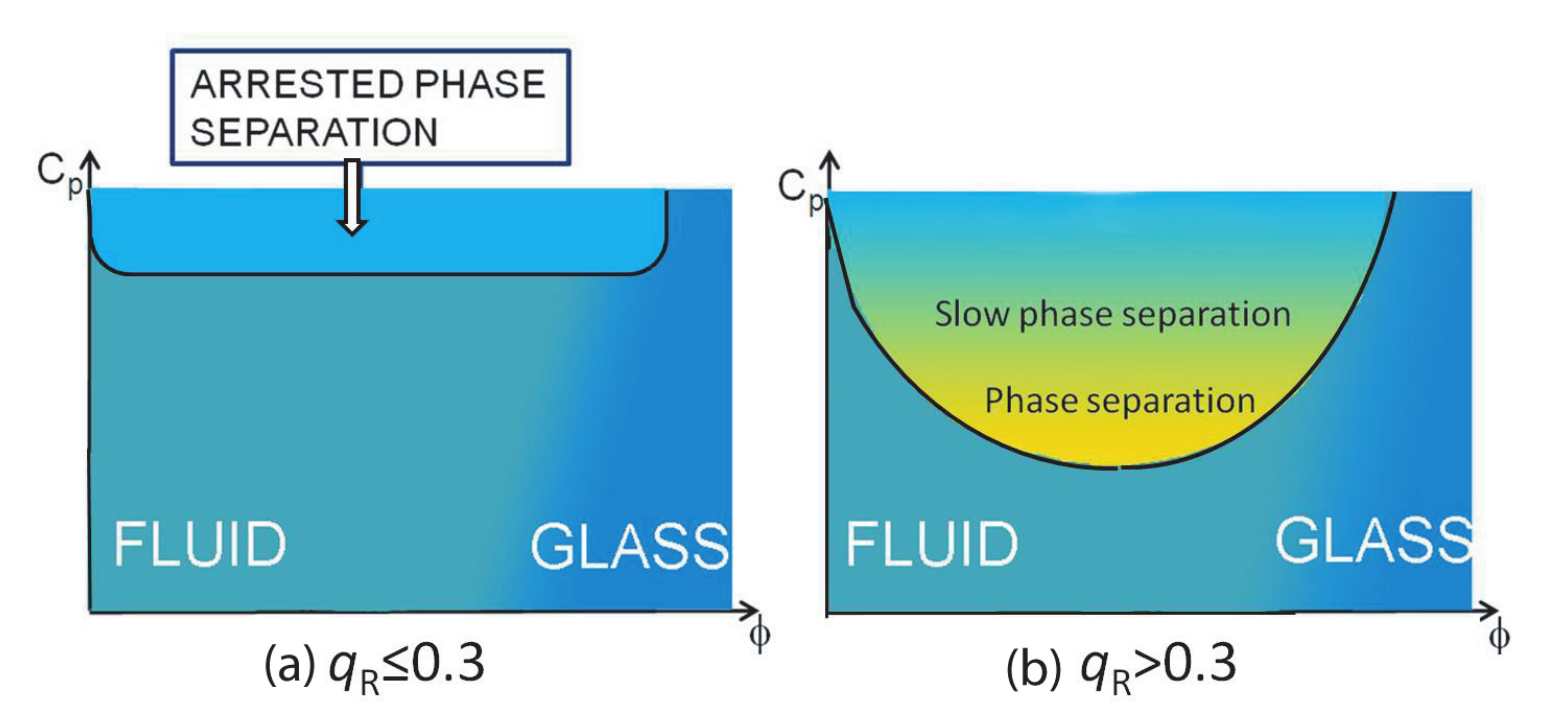}
\caption {Schematic phase diagram showing the effect of attractive range on the gas-liquid coexistence line and the intersection between the coexistence line and the glassy region. In short-range systems (a), the liquid-gas binodal is quite flat and upon quenching, the colloid-rich phase rapidly approaches a regime of glassy dynamics \cite{elliot1999}. Conversely, in longer-range systems (b), the volume fraction of the colloid-rich phase increases more gradually upon quenching.}
\label{figPhaseDiagramSchematic}
\end{figure}

While the topology of the phase diagram is well-known, the case of metastable amorphous states presents some challenges to theory and simulation \cite{taylor2012}. Smaller polymers lead to a shorter interaction range $q_\textrm{R}$, so that upon demixing, the colloids are closer together and thus the volume fraction of the colloid-rich phase $\phi_\textrm{c-r}$ is increased \cite{elliot1999,foffi2002}. Deep quenches (higher polymer concentration) may also raise $\phi_\textrm{c-r}$. During phase separation, if $\phi_\textrm{c-r}$ reaches volume fractions at which dynamical arrest occurs ($\phi_\textrm{c-r} \sim 0.57-0.59$), demixing is suppressed and a gel results \cite{lu2008,zaccarelli2007,tanaka2005}. In the absence of crystallisation, such a system is nonetheless metastable to colloidal liquid-gas coexistence. Fig. \ref{figPhaseDiagramSchematic} shows a schematic sketch of the effect of interaction range on the phase behaviour (ignoring crystal phases) \cite{foffi2005,lekkerkerker1992}.

Although such polydisperse dense (colloidal) liquids may be metastable to the formation of size-segregated crystals \cite{sollich2010}, to the best of our knowledge this has not been observed in experiments. Consequently, while most theoretical treatments necessarily consider the equilibrium monodisperse case where colloidal liquids are found up to $\phi_\textrm{c-r} \lesssim 0.5$ (a regime in which dynamical arrest is not expected),  the local volume fraction in the `arms' of a gel may be higher in experiments due to polydispersity. Although mean-field theories exist where the metastable gas-liquid binodal and spinodal can be calculated \cite{schmidt2002}, such predictions have limited accuracy for many polymer-colloid size ratios \cite{taylor2012}.

Previous work has largely focused on gelation in systems with short-range attractions, typically with polymer-colloid size ratios $q_{\textrm{R}} \lesssim 0.2$\cite{lu2008,poon1995,segre2001,ramakrishnan2002,verhaegh1999, royall2008} such that there is no equilibrium colloidal liquid phase. In such systems, gelation occurs upon crossing the metastable fluid-fluid spinodal\cite{lu2008,manleyWyss2005,ramakrishnan2002,royall2008}.  In contrast, phase separation and its arrest in colloid-polymer mixtures with medium-to-long range attractions, where there is a stable colloidal liquid phase, are much less understood.
Recent simulation\cite{testard2011} and experimental data\cite{lietorSantos2008,teece2011} suggest that a liquid-like colloid-rich phase exists, which grows continuously in length scale as the system ages.

For long-range ($q_{\textrm{R}} \gtrsim 0.3$) systems, we suggest two regimes of interest in relation to the gas-liquid binodal - shallow and deep quenches. For shallow quenches, the kinetics of phase separation in colloid-polymer mixtures is analogous to liquid-liquid demixing in molecular systems \cite{bailey2007}. The structural evolution can be described by the Cahn-Hilliard theory of spinodal decomposition \cite{cahn1959, jones2002}.  Several regimes are distinguished by the growth rate of the wavelength of density fluctuations. Immediately after quenching, a fastest-growing dominant mode $q_{\textrm{m}}$ appears with a wavelength $\lambda_{\textrm{m}}$, which exhibits time-dependent growth, with the power-law exponent changing from $1/3$ to 1 as the system evolves from the diffusive into the hydrodynamic coarsening regime. The characteristic domain size associated with the wavelength grows until it approaches the capillary length $\lambda_{\textrm{c}}=(\gamma / \Delta\rho g)^{1/2}$, where $\gamma$ is the interfacial tension and $\Delta\rho$ the density difference between the two phases, at which point any gravity-driven flow may rupture the bicontinuous structure \cite{aarts2005}.  Previous work on spinodal decomposition in colloid-polymer mixtures has shown that the crossover between diffusive and hydrodynamic regimes is broad and continuous, and both theory and experiment indicate that, depending on the importance of hydrodynamic interactions, the exponent may vary between 0.2 and 1.1 \cite{dhont1996,bailey2007}.

Quenching more deeply, the volume fraction of the colloidal liquid increases \cite{lekkerkerker1992,schmidt2002}.  The volume fraction of the colloid-rich phase may then approach $\phi \sim 0.58$, the point at which slow dynamics set in for hard spheres. Coarsening may therefore be suppressed and the bicontinuous network generated by the initial spinodal decomposition becomes partially or, in the limit of very short-range systems, even completely arrested on the experimental timescale \cite{tanaka2005}. The viscosity difference between the phases may be reminiscent of the viscoelastic phase separation proposed by Tanaka \cite{tanaka2005}.



In this work we explore these phase separation regimes in colloid-polymer mixtures with intermediate-to-long range attractions ($0.45 \leq q_\textrm{R} \leq 0.89$) using confocal microscopy.  Our data suggest that there is a continuous transition from gas-liquid phase separation to arrested phase separation (gelation) as a function of polymer concentration. The phase separation dynamics become increasingly slow as the system is quenched more deeply.  Experimental phase diagrams show that as the attractive range increases, the region where phase separation is uninterrupted by gelation expands. In addition, the mechanism of domain growth is studied in detail. We find that upon quenching, coarsening occurs predominately through enhanced particle mobility at the surface of the strands of the gel network.

\section{Materials and methods}
\subsection{Colloid-polymer mixture I: Emulsions + HEC}

Our first model system consisted of a low polydispersity emulsion of poly(dimethyl siloxane) (Silicone DC200 fluid 10cSt, Serva) dispersed in a refractive index-matched mixed solvent of 1,2-ethane diol (EG) and water. The emulsion was stabilized against aggregation by a combination of three commercial non-ionic surfactants (w/w ratio 1:0.417:0.015); tristyrylphenol ethoxylate, 16EO (Tanatex), PEO-PPO-PEO, $M_\textrm{w}$ 4950, 30\% PEO (Pluronic P103, BASF) and to give a small negative charge, the anionic surfactant sodium bis(2-ethyl 1-hexyl) sulfosuccinate (Aerosol OT, Cytec).  Small droplets (PDMS:surfactant 1:0.12) had a hydrodynamic radius $\sigma/2=142$ nm and a size polydispersity of 0.10, while large droplets (PDMS:surfactant 1:0.05) had a hydrodynamic radius  $\sigma/2=277$ nm and a size polydispersity of 0.18.  The index-matched continuous phase had a viscosity of 5.42 mPa
s $\pm$ 0.02 mPa s, as measured by capillary viscometry, a refractive
index of 1.396 at 543 nm, and a density of $\rho_{\textrm{c}} =1.067$ g cm$^{-3} \pm$
0.005 g cm$^{-3}$. The density of the PDMS oil is lower than the
continuous phase ($\rho_{\textrm{m}} = 0.934$ g cm$^{-3} \pm 0.006$ g cm$^{-3}$) so that the dispersions cream in a gravitational field. The density mismatch
of the oil is $\Delta\rho = \rho_\textrm{c} - \rho_\textrm{m} = 0.13\;\pm\;$ 0.01 g cm$^{-3}$. To screen out long-range electrostatic interactions and generate hard-sphere-like interactions between the emulsion drops we add 3 mM potassium chloride, which gives a Debye length of 6 nm.  A dye, Sudan Black, was also added to increase contrast between colloid-rich and colloid-poor phases such that the phase separation could be identified by eye.  Sudan Black preferentially adsorbs to the surface of the PDMS droplets, and so the colloid-rich phase appears dark and the colloid-poor phase light.

Hydroxyethylcellulose (HEC), which is a water-soluble non-ionic cellulose derivative, was used as the depletant.  Two molecular weights, of hydroxyethylcellulose were employed: Natrosol 250 M and Natrosol 250 HHX (Ashland-Aqualon), with molecular weights of $M_\textrm{w}=7.2\times10^{5}$ g mol$^{-1}$ and $M_\textrm{w}=1.3\times10^{6}$ g mol$^{-1}$ respectively.  Their estimated radii of gyration $R_{\textrm{g}}$ are 92 nm for Natrosol 250 M and 126 nm for Natrosol 250 HHX, as obtained by static light scattering.  Small droplets combined with Natrosol HHX gives a polymer-colloid size ratio, $q_{\textrm{R}} = (2 R_{\textrm{g}} / \sigma)$, of 0.89; small droplets with Natrosol M gives $q_{\textrm{R}}=0.65$; finally, large droplets with Natrosol HHX gives $q_{\textrm{R}}=0.45$.

Samples were prepared by first diluting a stock solution of the polymer with solvent, ensuring proper dispersion of the polymer. Then, solvent containing fluorescent rhodamine-B dye (Sigma Aldrich) was added such that the concentration of rhodamine-B was 0.02 gL$^{-1}$ in the aqueous phase for all samples. Finally, stock emulsion was added and the sample gently stirred for at least ten minutes before being quickly transferred to the imaging cell.  The start of the experiment ($t = 0$) was taken as the instant when stirring ceased.

\subsection{Colloid-polymer mixture II: PMMA + PS}

For our second system, we used poly(methyl methacrylate) (PMMA) core-shell particles\cite{bosma2002,campbell2002}, sterically stabilised with poly(12-hydroxy stearic acid) (PHSA), with a mean radius 416 nm and a polydispersity of 0.10, as determined by scanning electron microscopy.  Both the core and shell were labeled with fluorescent dyes: the core with coumarin and the shell with rhodamine-B.  The particles were suspended in a density-matched 78/22 (\%w/w)
mixture of cyclohexyl bromide and cis-decalin. 4 mM of tetrabutyl ammonium bromide was dissolved in the solvent mixture to screen electrostatic interactions and ensure hard-sphere-like behaviour of the colloids.  Depletion attractions were induced by addition of polystyrene (PS) as a non-adsorbing polymer.  The polymer molecular weight was $2.06\times10^{7}$ g mol$^{-1}$.  We estimate the polymer radius of gyration, $R_\textrm{g}$, to be 203 nm\cite{vincent1990}, such that the polymer-colloid size ratio $q_{\textrm{R}}$ was 0.49, which was comparable to that of the smallest size ratio in the emulsion droplet system.

\subsection{Confocal microscopy}
Confocal microscopy was used to study the temporal evolution of the bicontinuous network formed in our aqueous emulsion samples. A supporting frame was constructed to allow a light microscope (Zeiss, Axioskop S100) fitted with a confocal scanning head (Zeiss, LSM Pascal) to be mounted horizontally.  A laser wavelength of 532~nm was used to excite rhodamine molecules dissolved in the solvent, such that the background appears light and the emulsion droplets dark.  For clarity, we have inverted the micrographs pertaining to system I so that the colloid-rich phase appears bright and the colloid-poor phase dark. Emulsion-polymer dispersions (system I) were studied in large rectangular glass cuvettes with the front wall of the cell constructed from a 170 $\mu$m thick optical quality cover glass. The sealed cuvette had a square cross-section with an internal dimension of 13 mm, a height of 30 mm, and contained a small hole at the top of the cell through which the cell was filled.

For PMMA samples (system II), a conventionally mounted confocal microscope (Leica SP5 with resonant scanner) was used to image in the $xy$-plane and a laser with wavelength 488~nm was used to excite fluorescence from the coumarin-labelled core of the particles. 2D Images were collected at time intervals of $\delta t=$150-300 milliseconds over a total time period of 10-20 seconds, with typically 100-400 particles in each frame.  Data was averaged over 10-20 runs in order to improve statistics.

\subsection{Analysis}
To study the structural evolution quantitatively, length scales characteristic of the gel structure were extracted from the 2D-confocal images.  For an image of width $L$  taken at time $t$, we calculated the radially-averaged structure factor $S(q,t)$

\begin{equation}\label{eqn:S(q)}
    S(q,t) = \frac{1}{2 \pi q \Delta q} \int_{q \leq |\mathbf{q'}| \leq q + \Delta q} \mathrm{d}\mathbf{q'} \left < \tilde{I}(\mathbf{q'},t) \tilde{I}(-\mathbf{q'},t) \right >
\end{equation}

\noindent
where $\tilde{I}(\mathbf{q},t)$ is the 2D Fourier Transform (FT) of the image intensity $I(\mathbf{r},t)$ and $\Delta q = 2 \pi /L$.  The 2D Fourier transform displays a ring of high intensity at low $q$ which signifies the presence of a large characteristic length scale.  We characterise the dominant wavevector by the first moment $\langle q(t) \rangle$ of the structure factor, where

\begin{equation}
\langle q(t) \rangle = \int_{a}^{b} \mathrm{d}q \; qS(q,t) / \int_{a}^{b} \mathrm{d}q \;S(q,t)
\end{equation}

\noindent
with limits $a=0.02$ $\mu$m$^{-1}$ and $b=0.4$ $\mu$m$^{-1}$.  The dominant wavevector was then converted into a characteristic size $\lambda_{\textrm{c}}(t)=2\pi / \langle q(t) \rangle$.

In the PMMA system, the motion of the colloidal particles was tracked using modified versions of standard tracking routines\cite{crocker1996,royall2003}. Particles were separated into two groups: we define surface particles as those with $n=1-4$ neighbours throughout the image set, and bulk particles as ones with $n>4$.

In order for timescales to be compared directly in systems with different particle sizes and solvent viscosities, we quote all times in units of the characteristic Brownian time $\tau_{\textrm{B}}=3 \pi\eta_\textrm{L}\sigma^{3}/(4k_{\textrm{B}}T)$,  which is comparable to the time taken for a particle
to diffuse its own radius in a solvent of viscosity $\eta_\textrm{L}$.  For the PDMS emulsion system, the long time viscosity $\eta_\textrm{L}$ of the continuous phase was determined from  the diffusion constant of particles of known radius suspended in HEC solutions, using dynamic light scattering.  Viscosities $\eta_\textrm{L}$ were deduced from the measured diffusion constant using the Stokes-Einstein equation $D=k_\textrm{B}T/(3\pi\eta_\textrm{L} \sigma)$.  The concentration dependence of the viscosity was then fitted to the Martin equation \cite{teece2011Arxiv}, $\eta_\textrm{L}/\eta_{0}=1+[\eta]\cp\exp(k_{\textrm{H}}[\eta]\cp)$, where $\eta_\textrm{L}$ is the limiting low shear viscosity, $\eta_{0}$ is the solvent viscosity, $[\eta]$ is the intrinsic viscosity and $k_{\textrm{H}}$ is the Huggins constant. This equation is a reasonable prediction of polymer viscosity in the semi-dilute regime.  Regression gave fitted parameters of $[\eta]=0.30$ Lg$^{-1}$ and $k_{\textrm{H}}=1.2$ for the HHX grade, and $[\eta]=0.25$ Lg$^{-1}$, and $k_{\textrm{H}}=1.2$ for the lower molecular weight M grade.   The estimated Brownian times $\tau_{\textrm{B}}$ were between 0.1 s and 1.1 s, depending on the particle size, polymer molecular weight and polymer concentration.  In the PMMA system, the Brownian time was determined as $\tau_\textrm{B} = 0.7$ s.

\subsection{Phase diagrams}

\begin{figure*}[!htb]
\centering
\includegraphics[width=160mm]{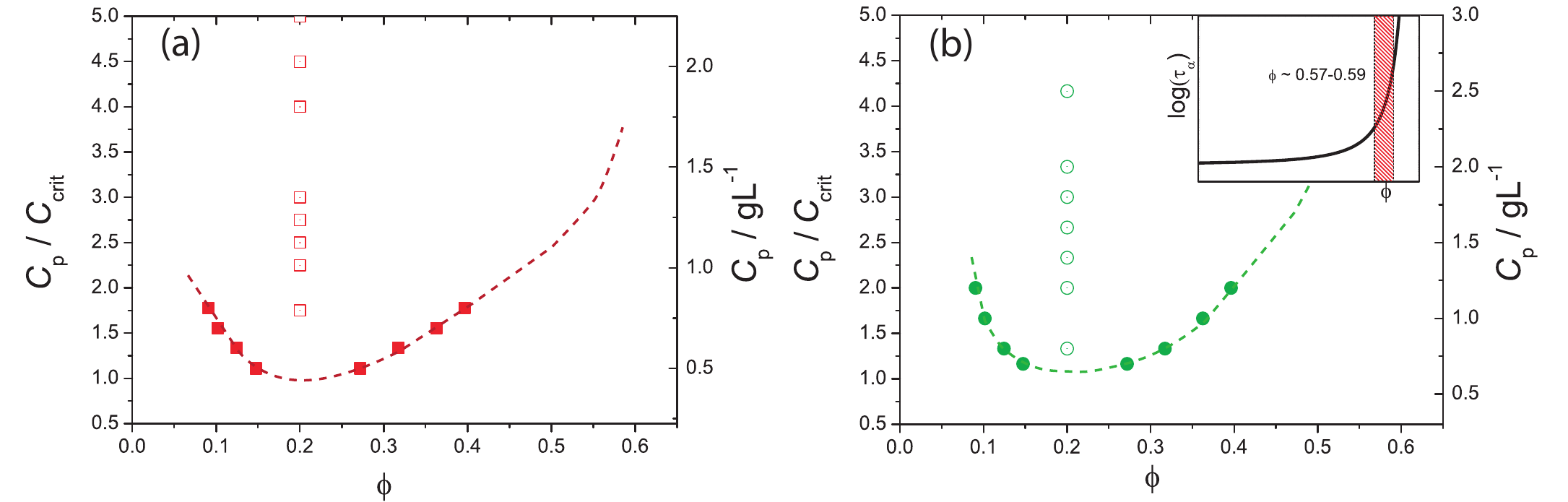}
\caption{Phase diagrams at constant overall sample colloidal volume fraction $\phi=0.2$, (a)$q_{\textrm{R}}=0.45$ and (b)$q_{\textrm{R}}=0.89$.  In each phase diagram, the left-hand $y$-axis gives polymer concentrations rescaled in terms of the critical polymer concentration $C_{\textrm{crit}}$ as estimated from the phase diagram, and the right-hand $y$-axis gives polymer concentration in units of gL$^{-1}$.  Dotted lines show qualitative extrapolations of the phase boundary for deeper quenches.  Filled-in data points indicate data from phase separation experiments, unfilled data points show state points used for confocal imaging and Fourier lengthscale data.  Inset shows dependence of structural relaxation time $\tau_{\alpha}$ on colloid volume density for hard spheres. Shaded region in inset denotes the range of $\phi=0.57-0.59$ where the dynamics become very slow.
\label{figUVVISPhaseDiagram}}
\end{figure*}

Experimental phase diagrams at size ratios $q_{\textrm{R}}=0.45$ and $q_{\textrm{R}}=0.89$ were constructed for system I.  The overall colloid volume fraction in each sample was fixed at $\phi=0.2$ for all samples, and polymer concentration $\cp$ and hence quench depth was varied, such that all state points lie on a vertical line through the phase diagram.  Phase diagrams were obtained by allowing emulsion droplet samples shallowly quenched into the two-phase region to fully equilibrate, then measuring the colloid volume fraction in both phases using UV-visible spectroscopy to measure absorbance of Sudan Black at 600 nm, which was proportional to the droplet concentration. Fig. \ref{figUVVISPhaseDiagram} shows the phase diagrams we obtained. Note that we fix the overall $\phi$ in each sample. Thus our representation is approximately analogous to the polymer reservoir representation  \cite{lekkerkerker1992}, scaled by the free volume fraction for $\phi=0.2$.

The gas-liquid coexistence boundaries at $\phi=0.2$ were at $\cp=0.40$ gL$^{-1}$ and $\cp=0.60$ gL$^{-1}$, for $q_{\textrm{R}}=0.45$ and $q_{\textrm{R}}=0.89$ respectively.  As the time taken to equilibrate increases rapidly with polymer concentration, only shallow quenches were experimentally accessible for this method and we note that underestimation of the colloidal volume fraction of the colloid-rich phase cannot be ruled out.  Although the polymer concentrations used here are less than those used for the confocal experiments, extrapolations are sketched to show that the in the case of the smaller $q$, colloid-rich phase becomes dense with relatively shallow quenches after the coexistence line is crossed.  In contrast, the dense branch of the long-range system requires much deeper quenches to achieve comparable volume fraction.

Confocal microscopy was used to identify state points as either one- or two-phase and therefore to determine an estimate of the phase boundary for $q_{\textrm{R}}=0.65$ (system I) and also for system II. The phase boundary for $q_{\textrm{R}}=0.65$ was estimated as $\cp=1.00 \pm 0.20$ gL$^{-1}$ at $\phi=0.2$. Now the phase boundary for $q_{\textrm{R}}=0.65$ in units of gL$^{-1}$ does not lie between those corresponding $q_{\textrm{R}}=0.45$ and $q_{\textrm{R}}=0.89$. This is due to the different polymer molecular weights used to obtain these size ratios.
Similarly for system II, the phase boundary was estimated as $\cp=0.13 \pm 0.03$ gL$^{-1}$.

As the molecular weights and the solvency conditions of the polymers varied depending on $q_{\textrm{R}}$ and also on the system used, polymer concentrations are scaled by the polymer concentration at the (colloidal) gas-liquid critical point  $C_{\textrm{crit}}$, so as to allow direct comparisons to be made.  We estimate $C_{\textrm{crit}}$ by taking the polymer concentration at the liquid-gas phase boundary at $\phi=0.2$. Since our phase diagrams (Fig. \ref{figUVVISPhaseDiagram}) are plotted analogously to the polymer reservoir represenation, close to criticality the phase boundary is quite flat with respect to $\phi$ so that $\phi=0.2$ provides a reasonable estimate for the critical volume fraction $\phi_\textrm{crit}$ \cite{vink2005,royall2007}.  For system I, $C_{\textrm{crit}}$ = 0.40 gL$^{-1}$, 1.00 gL$^{-1}$ and 0.60 gL$^{-1}$ for $q_{\textrm{R}}=0.49$, $q_{\textrm{R}}=0.6$ and $q_{\textrm{R}}=0.89$ respectively.  For system II, $C_{\textrm{crit}}=0.13$ gL$^{-1}$.

\section{Results and discussion}

\subsection{Time evolution: observation}
At short times ($>$1 minute), spinodal decomposition patterns were observed. For very shallow quenches, in the case of system I (not density-matched), instead of gel formation, gravity-driven flow occurs almost immediately after a bicontinuous network forms, reminiscent of images showing the Rayleigh-Taylor instability in colloidal suspensions\cite{aarts2005,wysocki2010}.  Deeper quenches allow a stress-supporting network to form, where gel-like structures evolved $\sim$30 seconds after the sample mixing is ceased.  Shorter-range, strongly attractive samples produced a long-lived network of fine strands, whereas samples with longer-range attractions showed a continuously changing viscoelastic gel structure.

\begin{figure}[!htb]
\centering
\includegraphics[width=80mm]{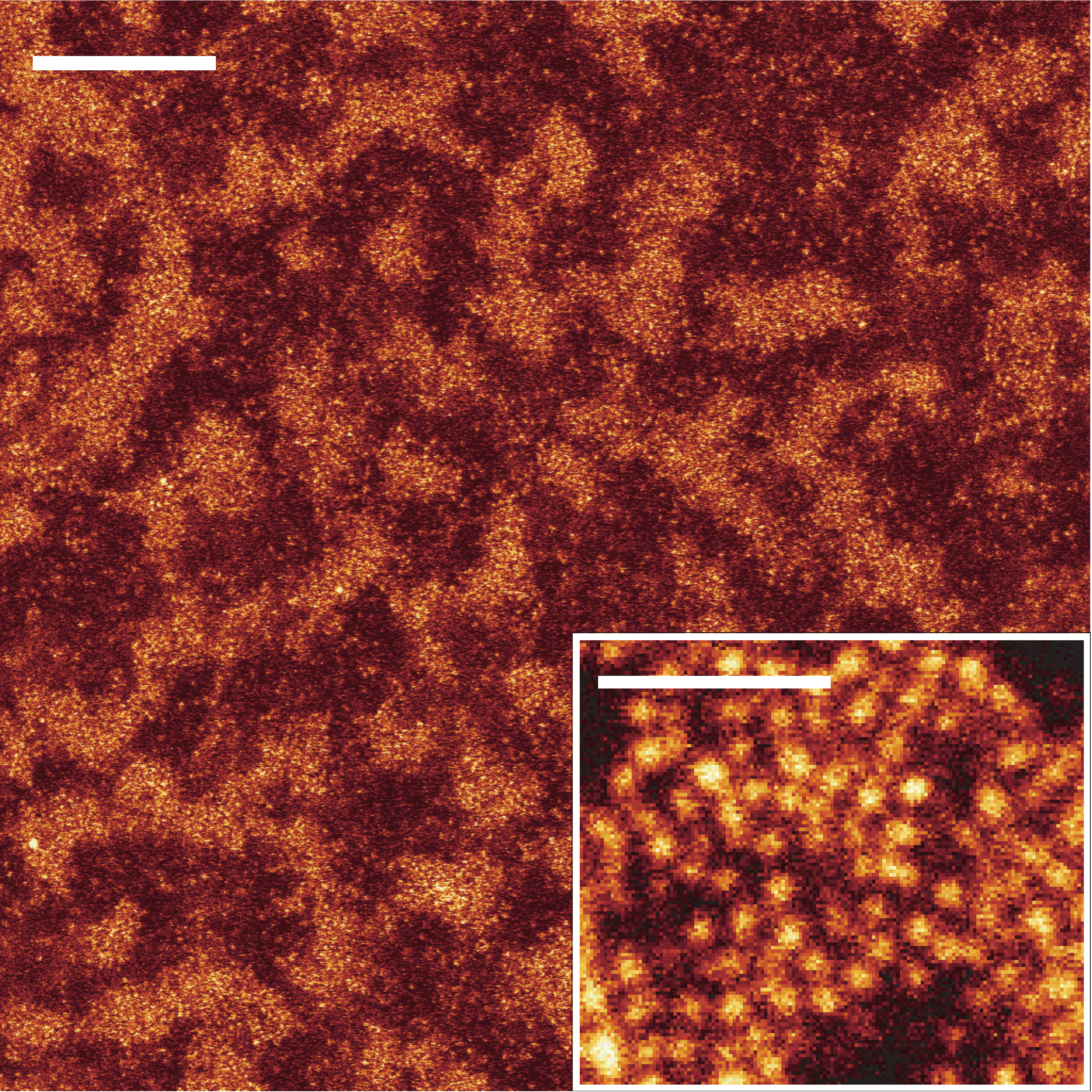}
\caption{PMMA particles undergoing fluid-fluid phase separation, $\cp/C_{\textrm{crit}}=1.20$, $\phi=0.2$, $q_{\textrm{R}}=0.49$ showing characteristic spinodal pattern, with inset showing detail.  Scale bar in top left-hand corner indicates 25 $\mu$m; scale bar in inset indicates 5 $\mu$m. }
\label{figSpinodalZoomout}
\end{figure}

At a polymer concentration of $\cp/C_{\textrm{crit}}=1.20$ (Fig. \ref{figSpinodalZoomout}) in the density-matched  system II, liquid-gas phase separation progresses to completion over about two hours, with a percolating network of the colloid-rich phase coexisting with a colloidal gas, highly reminiscent of spinodal-like patterns also seen in the emulsion droplet system (I).

\begin{figure}[!htb]
\includegraphics[width=80mm]{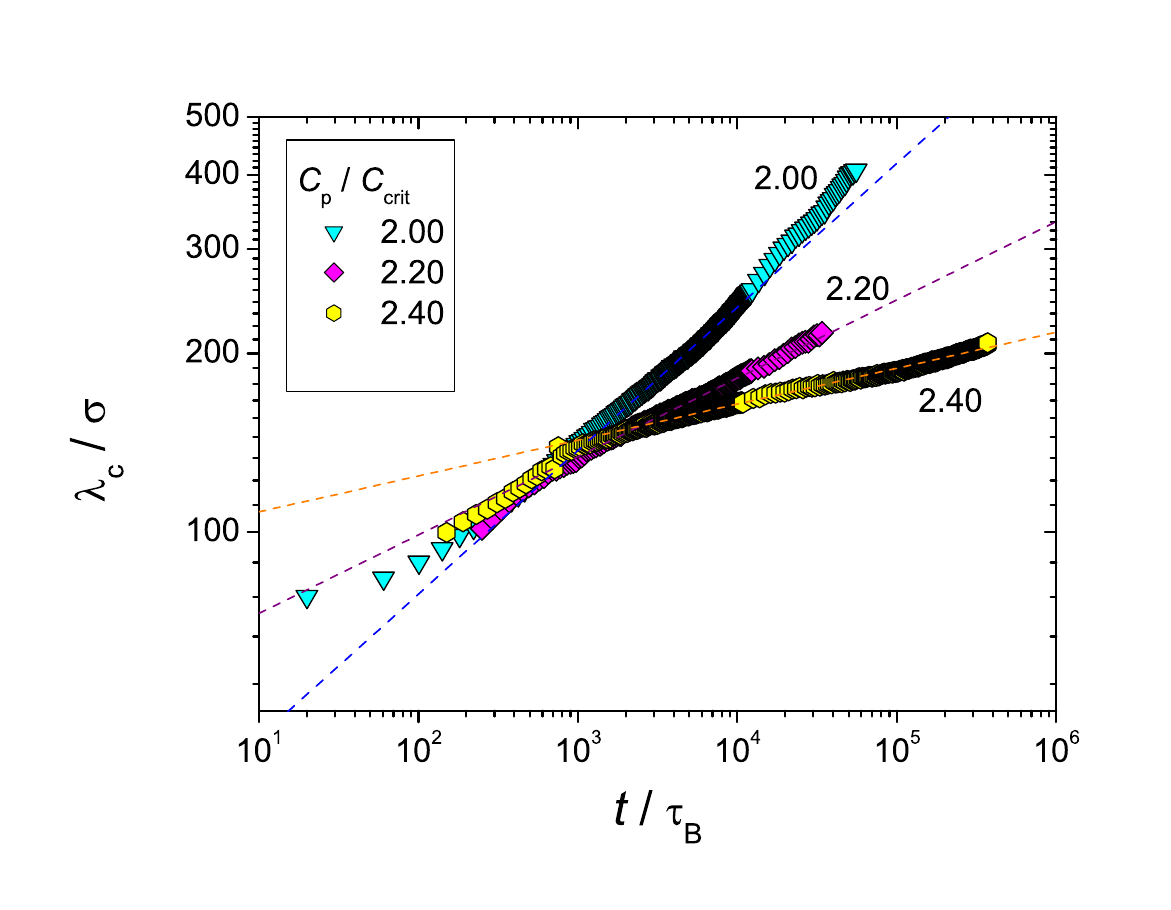}
\caption{Examples of growth exponent fits for selected domain size growth datasets.
Here we plot the characteristic domain size $\lambda_\textrm{c}$ as a function of time.
$q_{\textrm{R}}=0.65$, $\phi=0.2$. Growth exponents of 0.229$\pm0.011$, 0.132$\pm0.006$, 0.060$\pm0.004$ respectively.}
\label{figFitting}
\end{figure}

The rate of structural coarsening in system I can be represented by the growth of the characteristic domain size, as defined by the dominant wavevector of the 2D Fourier Transform of each confocal image.  Each data set exhibited a period of power law growth, where the rate of growth was given by the power law exponent.  Fig. \ref{figFitting} shows examples of how growth exponents were determined.

Data deviated from power law growth at very short times ($<100\tau_\textrm{B}$), where the domain size remained nearly constant.  This is indicative of a linear Cahn regime in early-stage spinodal decomposition and has previously been observed in other colloidal systems \cite{bhat2006,teece2011}.

The time-dependence of domain size $\lambda_\textrm{c}$ for varying polymer concentrations and size ratios is shown in Fig. \ref{figLengthscale}.  Growth exponents as a function of polymer concentration are shown in Fig. \ref{figGradients}.  At low polymer concentration, a regime of fluid-fluid phase separation was observed for all size ratios.  In this regime, the growth exponent is independent of quench depth for a given size ratio but ranged between 0.2 and 0.6 for the range of $q_{\textrm{R}}$ studied. This is consistent with previous work on phase separation in colloid-polymer mixtures where the growth exponent varied between 0.2 and 1.1 due to the broad crossover between diffusive and hydrodynamic regimes\cite{dhont1996,bailey2007}.

\subsection{Time evolution: connection to phase behaviour}
\label{subsectionTimeEvolution}
To gain insight into the phase separation dynamics, we assume these are related to the structural relaxation time in the colloid-rich phase. Neglecting the interfaces resulting from the bicontinuous network, the colloid-rich phase may be approximated by hard spheres at the same volume fraction, as the polymer concentration in this phase is low \cite{lekkerkerker1992}.  The structural relaxation time $\tau_{\alpha}$ in the colloid-rich phase can then be estimated, as a function of the volume fraction $\phi$, by the Vogel-Fulcher-Tammann equation\cite{brambilla2009, masri2009, flenner2010}:

\begin{equation}
\tau_{\alpha}(\phi) = \tau_{\infty} \exp\left[\frac{A}{(\phi_{0}-\phi)^{\delta}} \right]
\label{tau_alpha}
\end{equation}

\noindent Equation \ref{tau_alpha} is plotted in the inset in Fig. \ref{figUVVISPhaseDiagram} using parameters from ref. \cite{brambilla2009}, namely $A=2.25\times10^{-2}, \delta=2.0$ and $\phi_0=0.637$. As depicted in the inset to Fig. \ref{figUVVISPhaseDiagram}(b), when $\phi \sim 0.57-0.59$, relaxation times increase very strongly.

\begin{figure}[!htb]
\centering
\includegraphics[width=80mm]{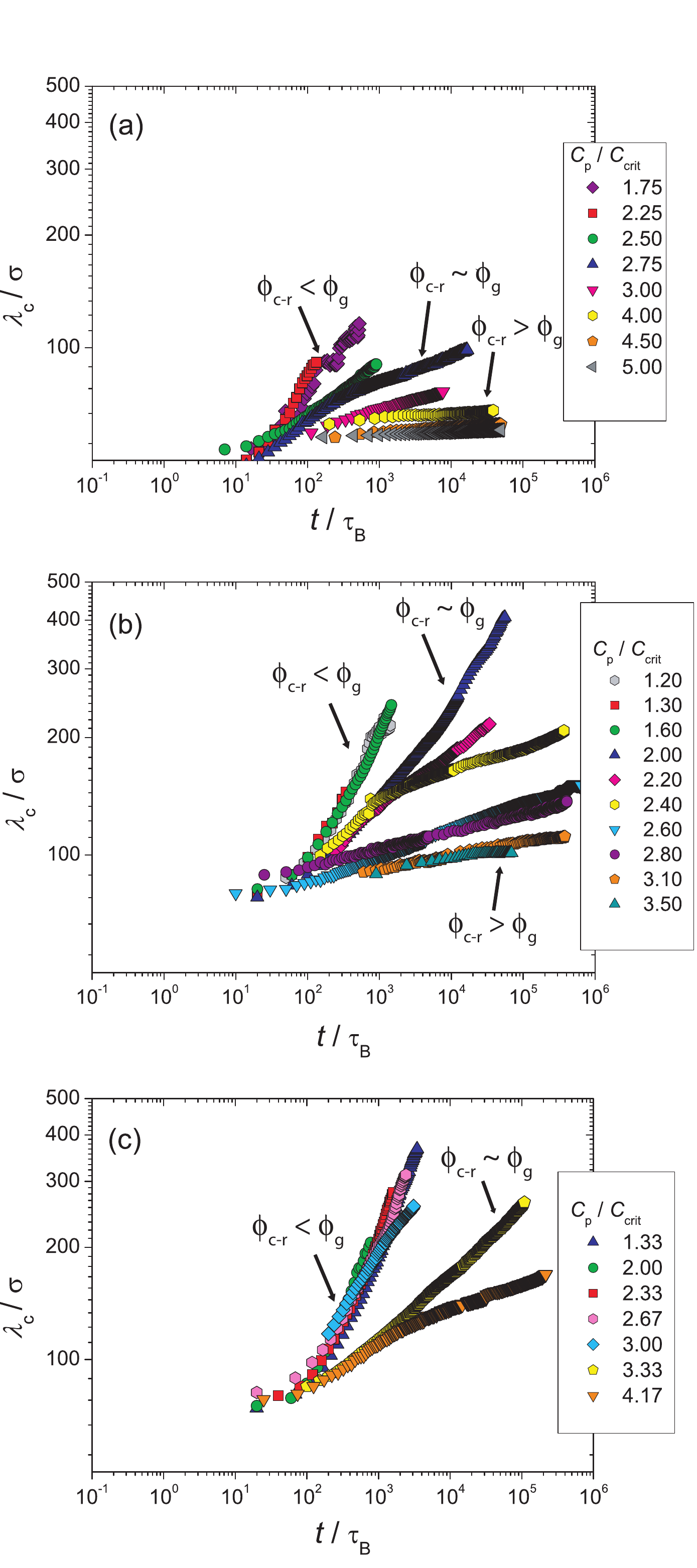}
\caption {Here we plot the characteristic domain size $\lambda_\textrm{c}$, obtained from 2D-FFT of confocal images of system I, as a function of time. $\phi=0.2$, (a) $q_{\textrm{R}}=0.45$, (b) $q_{\textrm{R}}=0.65$, (c) $q_{\textrm{R}}=0.89$.  Indicated in (a-c) are estimated volume fractions in the colloid-rich phase $\phi_{\textrm{c-r}}$, as compared volume fraction at which hard spheres undergo the glass transition, $\phi_{\textrm{g}}$.
}
\label{figLengthscale}
\end{figure}

The dashed lines in Figs. \ref{figUVVISPhaseDiagram} (a) and (b) indicate the expected increase in volume fraction of the colloid-rich phase $\phi_\textrm{c-r}$ at the state points considered in our time-evolution (Fig. \ref{figLengthscale}). Our extrapolation suggests that, in the intermediate and slow phase separation regimes, the volume fraction of the colloid-rich phase reaches $\phi_\textrm{c-r}\sim0.57-0.59$, such that the dynamics of phase separation is limited by particle relaxation in the dense phase. In other words, the particle level dynamics become coupled to phase separation as a whole. Conversely, at weaker quenches where the volume fraction of the colloid-rich phase is not high enough to exhibit slow dynamics, $\tau_{\alpha}$ is very much less than the dynamics of phase separation. In short, our system exhibits two regimes of phase separation - `normal' and `viscoelastic'\cite{tanaka2000}.

\begin{figure}[!htb]
\centering
\includegraphics[width=80mm]{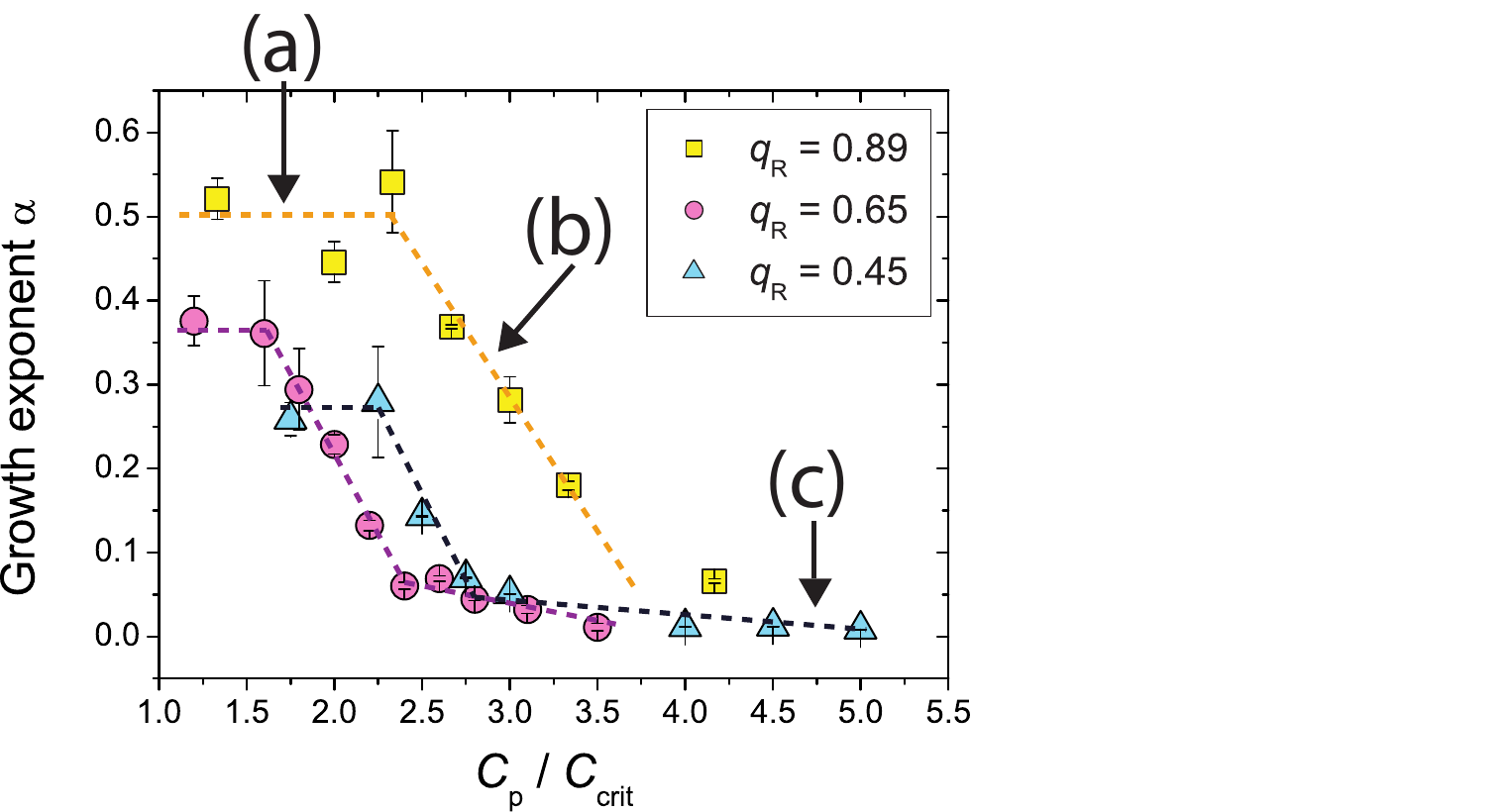}
\caption{Growth exponent of domain sizes as a function of polymer concentration, $q_{\textrm{R}}=0.45 - 0.89$, $\phi=0.2$.  Dashed lines are a guide to the eye. Labels indicate behavioural regimes as follows: (a) Normal phase separation, (b)Intermediate 'slow' phase separation and (c) gelation.}
\label{figGradients}
\end{figure}

This crossover between phase separation and gelation gives rise to an intermediate regime where phase separation still progresses to completion, but in contrast to ordinary fluid-fluid phase separation, the growth exponent decreases with increasing polymer concentration.  This regime is dominated by slow dynamics in the colloid-rich phase, where increasing densities result in a viscous yet still fluid state.  A corresponding increase in $\tau_{\alpha}$ can be seen, although in this regime $\tau_{\alpha}$ is still easily accessible on the experimental timescale.  These regimes are indicated in Fig. \ref{figGradients}.

The onset of gelation can be identified from the data by a decrease in the growth exponent during the imaging time, at $t\sim100$ seconds (which corresponds to around 1000$\tau_\textrm{B}$) as shown in Fig. \ref{figLengthscale}.  In these samples, phase separation is suppressed by gelation after the colloid-rich phase densifies to form a glass.  This type of behaviour has also been seen in simulations\cite{testard2011} of the Lennard-Jones system, as well as experiments on systems with short-range attractions\cite{lu2008}.  Quenching more deeply, the slow growth regime of characteristic domain size proceeds is quickly reaching after quenching and so the change in growth exponent becomes too fast to be experimentally accessible.

In the gel regime, the structure continues to coarsen, i.e. the growth exponent is non-zero.  This differs markedly from the behaviour of gels with short-range attractions and, at first glance, may appear paradoxical, as the gel structure is composed of a glassy network where particles are unable to rearrange.

\subsection{Local dynamics}

We now address this apparent paradox by studying the motion of individual particles using
 experimental system II. We find that that particle mobility at the surface of the gel strand is enhanced in comparison to particles in the interior of the gel strand, such that occasional particle-hopping was observed at the surface of the gel strands even in gels that were `arrested'.

\begin{figure}[h]
\centering
\includegraphics[width=80mm]{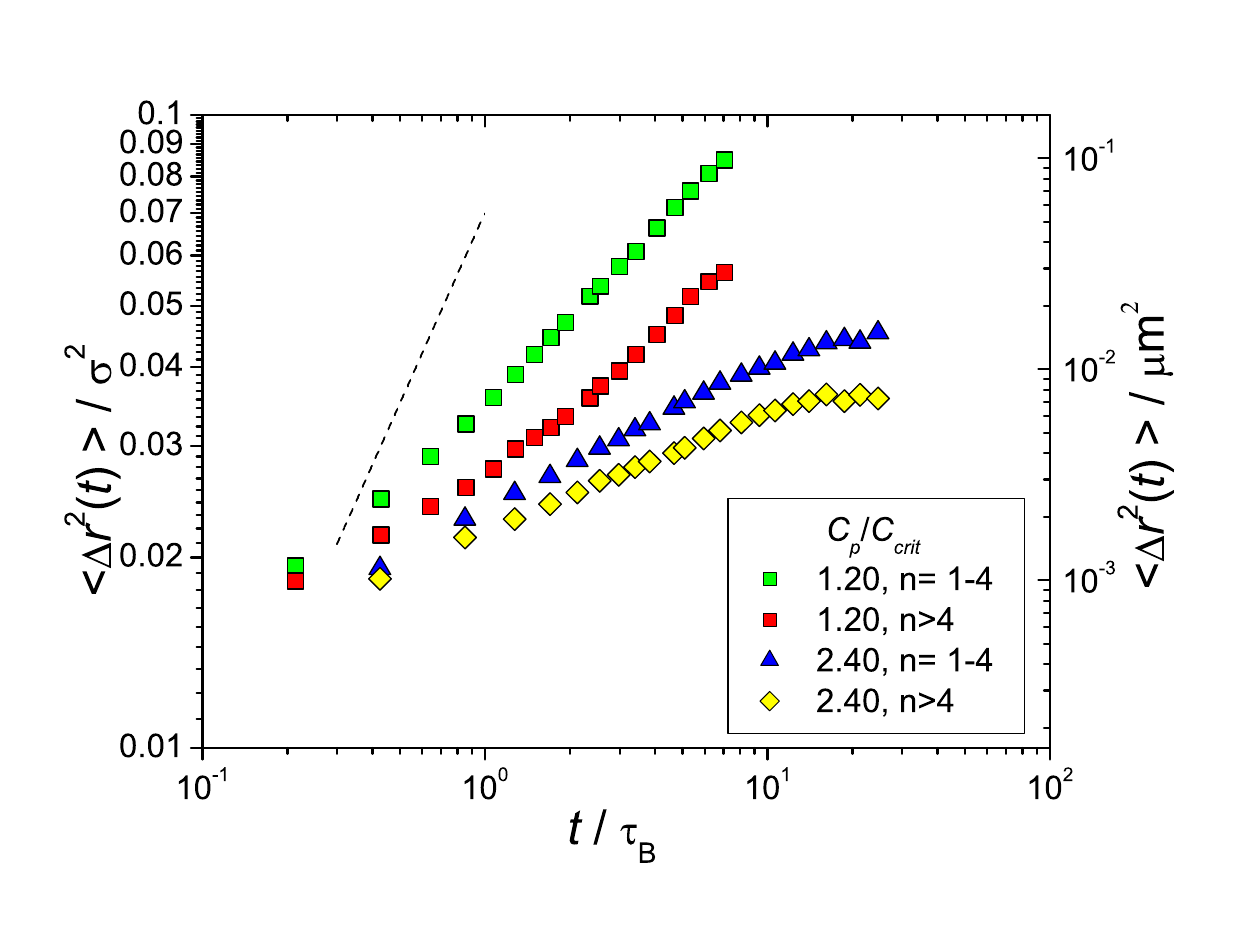}
\caption{Mean square displacements for $\cp/C_{\textrm{crit}}=1.20$ and $\cp/C_{\textrm{crit}}=2.40$ in units of $\mu$m$^{2}$ and particle diameter $\sigma$, with particles grouped according to number of neighbours $n$. Dotted line shows diffusive linear growth of MSD.  $\phi=0.2$, $q_{\textrm{R}}=0.49$}
\label{figMSD}
\end{figure}

Dynamical heterogeneities became significant as the system approached gelation.  At a polymer concentration of $\cp/C_{\textrm{crit}}=2.40$, it was apparent that the system consisted of two particle populations, an effect previously observed in colloidal systems\cite{gao2007,puertas2004,royall2008}, and also for polymer glasses\cite{fakhraai2008}.  `Slow' particles were stuck in an arrested network, and `fast' particles, which were able to diffuse over a number of particle diameters, occupied surface sites on this network.  Mean square displacements of bulk and surface particles (Fig. \ref{figMSD}) show that they correspond to `slow' and `fast' particles respectively.  Sub-diffusive behaviour is observed for all particles, although this may be due to the relatively short imaging time.   Due to the experimental timescale, the extent of particle diffusion is limited, but there are indications that a plateau may be developing in the data for $\cp/C_{\textrm{crit}}=2.40$.  As quench depth increased, the number of `fast' particles decreased.  At a deep quench of $\cp/C_{\textrm{crit}}=9.70$, no fast particles were detected over a two-hour period.

\begin{figure}[h]
\centering
\includegraphics[width=80mm]{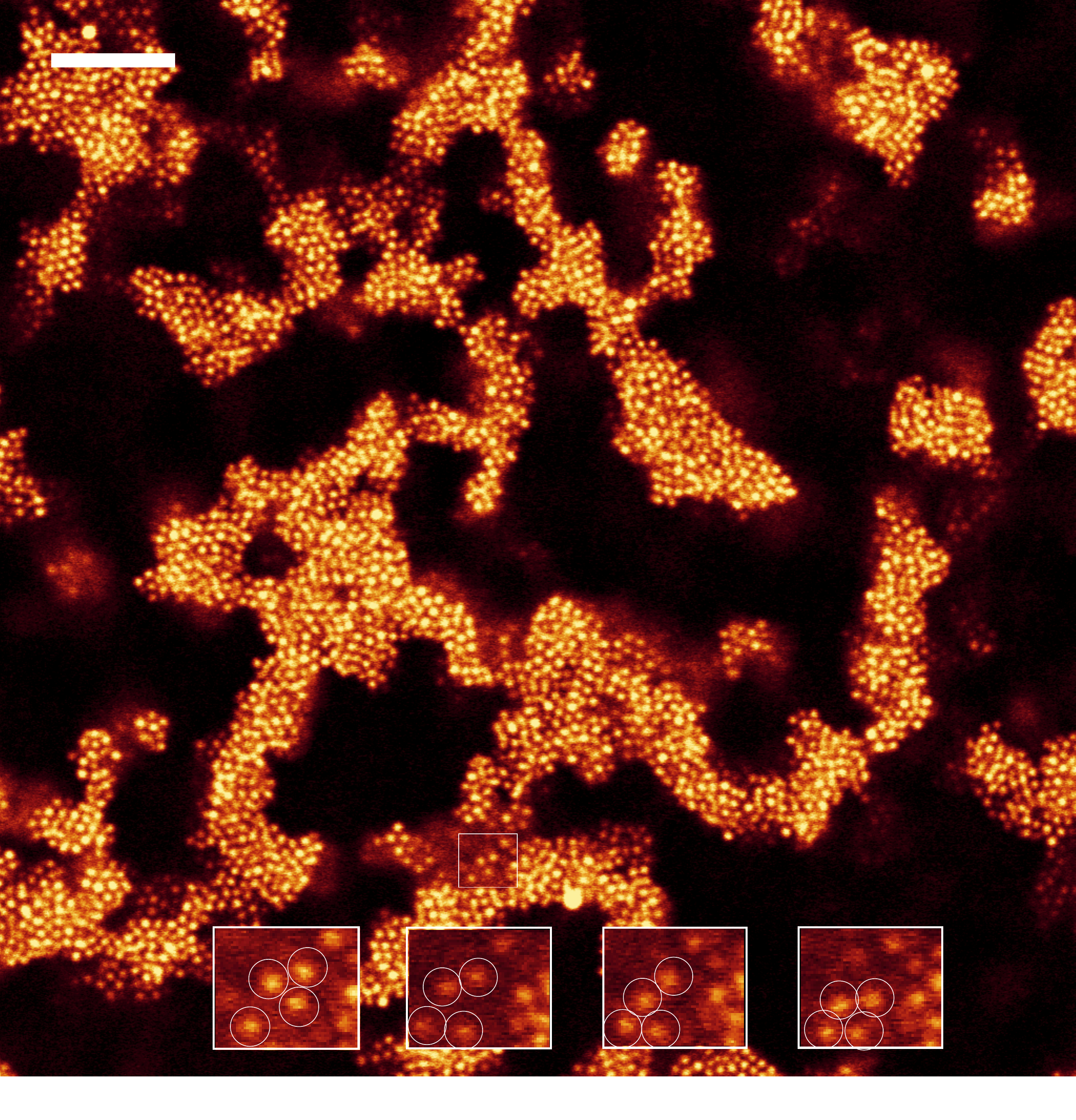}
\caption{Localised particle-hopping on surface of arrested gel network.  Magnified snapshots of an area (small box) are shown in the four larger boxes as a time-sequence.  The particles undergo a local re-arrangement over $\sim 500$ s, as indicated by the outlined particles. The snapshots are taken at $t=320$ s, 410 s, 420 s and 460 s after imaging began, which correspond to $460\tau_{\textrm{B}}$, $590\tau_{\textrm{B}}$, $600\tau_{\textrm{B}}$ and $660\tau_{\textrm{B}}$.
$\cp/C_{\textrm{crit}}=3.90$, $q_{\textrm{R}}=0.49$, $\phi=0.2$. Scale bar in top left-hand corner indicates 10 $\mu$m.}
\label{figSingleParticle}
\end{figure}

Some localised areas of the gel network surface exhibited greater particle mobility than others, and particles appear to rearrange cooperatively, an example of which is shown in Fig. \ref{figSingleParticle}.  It is possible that self-generated mechanical stress within the gel and the resulting deformation lowers the activation energy of particle-hopping in localised areas.  It has previously been proposed \cite{koyama2009} that self-generated stress builds up at local weak points within a transient gel network, thereby increasing particle mobility in small regions and providing a mechanism for gel collapse in arrested gels.

\section{Conclusions}

We have presented an experimental study of dynamics in phase-separating colloid-polymer mixtures with interaction range between 0.45 and 0.89.  Our data suggest that the crossover from phase separation to gelation is a continuous process as function of polymer concentration, with phase separation dynamics slowing gradually until arrest is reached.  By analysing the time growth of the domain size during phase separation, we observe a range of behaviour from liquid-gas phase separation at shallow quench depths to dynamic arrest and gelation at deep quenches.  Data from confocal imaging is supported by experimental equilibrium phase diagrams, in order to highlight the changes in dynamics with attractive range.  As the attractive range increases, the region of the phase diagram corresponding to phase separation becomes larger, and the crossover into gelation more gradual.

In addition, we find evidence of a mechanism of structural coarsening and domain growth in `arrested' gels, where rearrangements on the surface of the gel strand may continue despite a glassy, arrested state inside the gel strand, an effect which becomes more significant with increasing attractive range.

The dynamics of the colloid-rich phase are taken to be similar to hard spheres at the same estimated volume fraction. Under this assumption we find the dynamics of phase separation couple to the structural relaxation at the particle level in the colloid-rich phase upon deep quenches where slow dynamics set in. At weaker quenches, the phase separation dynamics decouple from the single-particle dynamics, and in this way we distinguish two modes of phase separation - `normal' and `viscoelastic'.  Further work could investigate this observation more closely.  In particular, a more accurate determination of the volume fraction of the colloid-rich phase would be desirable. While this may be tackled experimentally or by simulation, theoretical extensions of metastable liquid binodals and spinodals would be most helpful. While mean-field theory \cite{schmidt2002} is readily available for the range of volume fractions in the colloid-rich phase relevant to gelation ($\phi_\textrm{c-r}\gtrsim0.57-0.59$), especially for smaller size ratios, its accuracy may be limited \cite{taylor2012}. We hope to have raised the issue that colloid-polymer mixtures (and other model systems such as DNA-coated colloids \cite{geerts2012})frequently become involved in metastable liquid states at high volume fraction, and that treatment of such states with theories more sophisticated than mean-field (such as hierarchical reference theory \cite{loVerso2006}) or
self-consistent Ornstein-Zernike approximation \cite{foffi2002} may prove invaluable if extended to higher colloid volume fraction for a variety of interction ranges.

\section*{Acknowledgements}

This work was supported jointly by Bayer CropScience and the UK Engineering and Physical Sciences Research Council through the award of an Industrial CASE award to IZ. CPR acknowledges the Royal Society for financial support and EPSRC grant code EP/H022333/1 for
provision of equipment used in this work




\footnotesize{

\bibliographystyle{rsc} 
\providecommand*{\mcitethebibliography}{\thebibliography}
\csname @ifundefined\endcsname{endmcitethebibliography}
{\let\endmcitethebibliography\endthebibliography}{}

}

\end{document}